# Degree of Polarization in Quantum Optics through 2nd –generalization of Intensity

Ravi S. Singh[1#] and Hari Prakash[2,†]

[1] Department of Physics, D. D. U. Gorakhpur University, Gorakhpur-273009 (U. P.), INDIA.
[2] Department of Physics, University of Allahabad, Allahabad-211002 (U. P.), INDIA.

ABSTRACT

Classical definition of degree of polarization (DOP) is expressed in quantum domain by replacing intensities through quantum mechanical average values of relevant number operators and is viewed as 1st –generalization of Intensity. This definition assigns inaccurately the unpolarized status to some typical optical fields, e.g., amplitude-coherent phase-randomized and hidden-polarized light, which are not truly unpolarized light. The apparent paradoxical trait is circumvented by proposing a new definition of DOP in Quantum Optics through 2nd-generalization of Intensity. The correspondence of new DOP to usual DOP in Quantum Optics is established. It is seen that the two definitions disagree significantly for intense optical fields but coincides for weak light (thermal light) or for optical fields in which occupancy of photons in orthogonal mode is very feeble. Our proposed definition of DOP, similar to other proposals in literature, reveals an interesting feature that states of polarization of optical quantum fields depend upon the average photons (intensity) present therein.

Polarization of light, ensuring transversal character, is a centuries-old concept discovered by E. Bartholinus [1]. In Classical Optics, this trait of light is characterized by Stokes theory (parameters) [2] geometrically interpreted due to Poincare [3]. Remarkably, these Stokes parameters can also be applied to some optical quantum fields for inferring polarization nature, where they are defined to be quantum mechanical average values of the Stokes operators [4]. Although the polarization of optical field has acquired indispensible candidacy for demonstrating fundamental issues of Quantum Mechanics as well as performing ingenious experiments in Quantum Optics [5], yet basic understanding of optical-polarization in terms of spatio-temporal variables of optical fields remains unexplored.

Although the studies on optical-polarization may, largely, be classified in two extremes: perfect (complete) polarized state and unpolarized state, optical fields may exist in infinitely many states which are neither polarized nor unpolarized. In 1971, the unpolarized optical field is rigorously investigated wherein the structure of its density operator is discovered [6]. Other prominent works [7] on the state of unpolarized light have brought in new insights about its quantum nature in conjunction with its tomography. Also, in Ref. [6], it is emphasized that Stokes parameters prescribes insufficient condition for characterizing the state of unpolarized electromagnetic radiation, especially, when higher-order correlations between optical field-amplitudes are critical [8]. On the other hand, perfect polarized light is defined by requiring disappearance of light (signal) in at least one transverse orthogonal mode [9], although this treatment doesn't provide a procedure for testing whether an arbitrary quantum state of light is perfect polarized or unpolarized.

Modern approaches for ascertaining the state of polarization witnessed two complementary methods: computable-measures and operational-measures. The former - measure [10], based upon the 'notion of distance' of optical quantum states from the state of unpolarized light, is applied to introduce expressions for degree of polarization (DOP). On the other hand, the latter approach is nothing but Stokes-parametric approach. Notably, Klimov et al. [11] formulated a pragmatic and ingenious criterion for DOP in terms of minimal fluctuations in Stokes parameters on the Poincare sphere. Astute inspection of higher-order correlations in Stokes parameters / variables, where only equal numbers of Bosonic creation and annihilation operators are involved [12], buttresses clinching evidence against general propensity in favour of Stokes parameters because these Stokes-parametric correlation functions are, inherently, not synonymous to higher-order Glauber correlation functions [13]. Thus, not only the 'distance-based approach, being an abstract conception, lacks correspondence to classical description of the optical-polarization and transcribes variant values of DOP for the same quantum state but also skepticisms mount pertaining to 'operational-measures' due to unprecedented incisive analyses [14] highlighting inadequacy of the Stokes theory. Moreover, Luis [15] contrived, by drawing analogy from SU

(2) Lie algebra of Stokes operators to those for components of Jordon-Schwinger spin angular momenta [16], that SU(2) Q-function is most suitable distribution function for probabilistic description of optical-polarization of quantum states on Poincare sphere. SU (2) Q-function of quantum states is, in turn, applied to cast a DOP as a 'distance' from the uniform distribution possessed by unpolarized light. Later on, this SU (2) Q-function-approach is generalized to characterize the states of polarization of non-paraxial three dimensional optical field [17] of which description has witnessed various alternative approaches [18]. But, Karassiov [19] recognized that Stokes operators found distinct sort of Hilbert space for its operation in contrast to that of spin-angular momenta. This is why, recently in spirit of classical description of optical-polarization, quantum phase-space description of polarized optical field is carried out [20]. Nonetheless, some serious objections may be drawn to Luis-proposal: firstly, it does not ascribe the value unity for the coherent light (perfect polarized state), a multiphoton state, and secondly, SU (2) Q-function does not connect manifolds with different photon excitation.

Recent trends in Quantum Optics spearhead new physical effects such as quantum Darwinism, quantum imaging, ghost imaging and spatio-temporal multipartite entanglement [21] in which spatio-temporal features of optical field, its quantum state engineering and bases-based quantum measurements is harnessed. Vociferously, none of the preceding definitions of DOP, whether it may be a computable (distance-based) or operational (Stokes-parametric) or SU (2)-Q function, explore innate relationships possessed by spatio-temporal properties of the optical quantum fields.

Our viewpoint on optical-polarization stems from its classical-description, i.e., by the superposition of two transverse orthogonal harmonic oscillators of synchronized frequency emulating two transverse orthogonal components of a harmonic electromagnetic plane wave in any basis of description preserving non-random values of 'ratio of real amplitudes' and 'difference in phases' or non-random values of the 'ratio of their complex amplitudes' which defines the 'Index of Polarization' [22] for perfect polarized light. In quantum regime a quantum criterion is established [see Eq. (7) below] by invoking the fact, which is due to Mehta and Sharma [9]. This criterion not only prescribe a recipe for characterizing whether a light in any arbitrary quantum state is perfectly polarized and picking out simultaneously the 'characteristic-parameters' [23].

We urge that the vacuum state of optical field (virtual photons) must find a paramount position in theory of optical polarization. It is, therefore, in the present brief-report, we introduce an alternative judicious expression of DOP by employing $2^{nd}$-generalization of intensity in which virtual vacuum photons enter through projection operation. Our definition, contrary to other prevalent proposals for DOP, meets the very basic requirements of the term DOP in verbatim furnishing unit value for perfect polarized state (coherent state) and vanishing value for unpolarized state of light.

Firstly, we shall describe our criterion to characterize perfect optical-polarization to establish consistency. A plane monochromatic optical field propagating along z-direction in free space can, in general, be described by vector potential, $\vec{A}$ in the form,

$$\vec{A} = \hat{e}_x A_{0x} \cos(\psi - \phi_x) + \hat{e}_y A_{0y} \cos(\psi - \phi_y), \psi = \omega t - kz,$$

or in analytic-signal representation,

$$\vec{\mathcal{A}} = (\hat{e}_x \underline{A}_x + \hat{e}_y \underline{A}_y) e^{-i\psi}, \underline{A}_{x,y} = A_{0x,0y} e^{i\phi_{x,y}} \quad (1)$$

where $\underline{A}_{x,y}$ are classical complex amplitudes; $A_{0x,0y}$, real amplitudes and phase parameters, $\phi_{x,y}$ ($0 \leq \phi_{x,y} < 2\pi$) possess, in general, random spatio-temporal variation with angular frequency, $\omega$ in linear polarization basis $(\hat{e}_x, \hat{e}_y)$ of transverse plane to $\vec{k}$ $(= k\hat{e}_z)$ which is propagation vector of magnitude, k and $\hat{e}_{x,y,z}$ are unit vectors along respective x-, y-, z-axes forming right handed triad.

We consider a pellucid property of perfect polarized optical field, namely, the non-random ratio of complex amplitudes of transverse orthogonally polarized modes,

$$\underline{A}_y/\underline{A}_x = p, \tag{2}$$

as a definition. Here p is a non-random complex parameter in linear-polarization basis, $(\hat{e}_x, \hat{e}_y)$ and is termed as 'Index of Polarization' [22] which renders characteristic polarization parameters ('ratio of real amplitudes and difference in phases'). Evidently, one may note that polarized optical field (through non random, p) is effectively a mono - modal optical field since only one random complex amplitude suffices for its complete statistical description.

Additionally, if one introduces new parameters, $A_0$ (real random amplitude defining global intensity), $\chi_0$ (polar angle), $\Delta_0$ (azimuth angle), $\phi$ (random global phase) on a Poincare sphere, satisfying inequalities $0 \leq A_0$, $0 \leq \chi_0 \leq \pi$, $-\pi < \Delta_0 \leq \pi$, $0 \leq \phi < 2\pi$, respectively, involving transforming equations in terms of old parameters, $A_0 = (A_{0x}^2 + A_{0y}^2)^{1/2}$, $\chi_0 = 2\tan^{-1}(A_{0y}/A_{0x})$ and $\Delta_0 = \phi_y - \phi_x$; $\phi = (\phi_x + \phi_y)/2$, the analytical signal, $\vec{\mathcal{A}}$, in Eq. (1), yields a self-instructive form,

$$\vec{\mathcal{A}} = \hat{\varepsilon}_0 \mathcal{A}; \mathcal{A} = \underline{A}e^{-i\Psi}; \underline{A} = A_0 \, e^{i\phi},$$
$$\hat{\varepsilon}_0 = \hat{e}_x \cos\frac{\chi_0}{2} e^{-\Delta_0/2} + \hat{e}_y \sin\frac{\chi_0}{2} e^{\Delta_0/2}. \tag{3}$$

Interpretatively, this form of vector potential, $\vec{\mathcal{A}}$ in Eq. (3) may be construed as a single mode polarized optical field, statistically explicable by single complex amplitude, $\underline{A}$ polarized in the fixed direction, $\hat{\varepsilon}_0$ specifying the polarization mode, $(\hat{\varepsilon}_0, \vec{k})$. Here, the complex unit vector, $\hat{\varepsilon}_0$ ($\hat{\varepsilon}_0^* \cdot \hat{\varepsilon}_0 = 1$) assigns parameterized expression of 'Index of Polarization' on Poincare sphere, $p = \underline{A}_y/\underline{A}_x = \tan\frac{\chi_0}{2} e^{i\Delta_0}$. Visibly, the state of optical-polarization is specified by the non-random values of p, which, in turn, is fixed by non-random values of $\chi_0$ and $\Delta_0$ defining a point ($\hat{\varepsilon}_0$) on the unit Poincare sphere analogous to its counterpart Stokes parameters.

One may develop quantum theory for perfect optical-polarization on a similar classical lineage. In Quantum Optics the optical field, Eq. (1) is described by operatic-version of vector potential operator, $\vec{\mathcal{A}} = \left(\frac{2\pi}{\omega V}\right)^{1/2} [(\hat{e}_x \hat{a}_x + \hat{e}_y \hat{a}_y) e^{-i\psi} + \text{h.c.}] = \left(\frac{2\pi}{\omega V}\right)^{1/2} [(\hat{\varepsilon} \hat{a}_{\hat{\varepsilon}} + \hat{\varepsilon}_\perp \hat{a}_{\hat{\varepsilon}_\perp}) e^{-i\psi} + \text{h.c.}]$, in linear-polarization basis $(\hat{e}_x, \hat{e}_y)$ or in elliptic-polarization basis $(\hat{\varepsilon}, \hat{\varepsilon}_\perp)$ [24], respectively, where $\omega$ is angular frequency of the optical field and V is the quantization volume of the cavity, h.c. stands for Hermitian conjugate. Orthonormal properties of $\hat{\varepsilon} (= \varepsilon_x \hat{e}_x + \varepsilon_y \hat{e}_y)$ and $\hat{\varepsilon}_\perp (= \varepsilon_{\perp x} \hat{e}_x + \varepsilon_{\perp y} \hat{e}_y)$ provides the relationships between Bosonic-annihilation operators $\hat{a}_{\hat{\varepsilon}}$ ($\hat{a}_{\hat{\varepsilon}_\perp}$) with those in linear-polarization basis $(\hat{e}_x, \hat{e}_y)$ as,

$$\hat{a}_{\hat{\varepsilon}} = \varepsilon_x^* \hat{a}_x + \varepsilon_y^* \hat{a}_y, \quad \hat{a}_{\hat{\varepsilon}_\perp} = \varepsilon_{\perp x}^* \hat{a}_x + \varepsilon_{\perp y}^* \hat{a}_y. \tag{4}$$

The pure (mixed) dynamical state of a monochromatic optical beam, propagating along z-axis and polarized in the mode, $(\hat{\varepsilon}_0, \vec{k})$, may be specified by a state vector $|\psi\rangle$ ($\rho$) in its Hilbert space. Obviously, such light doesn't have signal (photons) in orthogonal mode $(\hat{\varepsilon}_{0\perp}, \vec{k})$, i.e.,

$$\hat{a}_{\hat{\varepsilon}_{0\perp}} |\psi\rangle = 0, \tag{5}$$

or, $\hat{a}_{\hat{\varepsilon}_{0\perp}} \rho = 0$, which yields, on applying Eq.(4), $(\varepsilon_{0\perp x}^{*} \hat{a}_x + \varepsilon_{0\perp y}^{*} \hat{a}_y)|\psi\rangle = 0$. Refurbishing it by orthogonality relation between $\hat{\varepsilon}_0$ and $\hat{\varepsilon}_{0\perp}$, one obtains the defining equation (criterion) for perfect optical-polarization,

$$\hat{a}_y |\psi\rangle = p\hat{a}_x|\psi\rangle, \tag{6}$$

or, $\hat{a}_y \rho = p\hat{a}_x \rho$, the quantum analogue to the classical perfect optical-polarization criterion, Eq. (2), giving $p = \varepsilon_{0y}/\varepsilon_{0x}$ $(= \tan\frac{\chi_0}{2} e^{i\Delta_0}$, from Eq. 3). Multiplying Eq. (5) from left by inverse annihilation operator, $\hat{a}_x^{-1} = \hat{a}_x^{\dagger}(1 + \hat{a}_x^{\dagger}\hat{a}_x)^{-1}$ [25], we obtain

$$\hat{P} |\psi\rangle = p(1 - \hat{V}_x)|\psi\rangle, \tag{7}$$

or, $\hat{P} \rho = p(1 - \hat{V}_x)\rho$, where $\hat{P} (\equiv \hat{a}_x^{-1}\hat{a}_y)$ is recognized as Polarization-operator and $\hat{V}_x$ is the vacuum projection operator for x-polarized virtual photons, $\hat{V}_x \equiv \sum_{n_y=0}^{\infty} |0, n_y\rangle\langle n_y, 0|$, $n_y$ is the number of y-polarized photons. As a demonstration of the criterion, Eq.(7) one may consider bi-photonic qutrit state, $|\psi\rangle = \frac{1}{3}|2, 0\rangle + \frac{2}{3}|1, 1\rangle + \frac{2}{3}|0, 2\rangle$ which provides $p = \sqrt{2}$ showing that light is plane-polarized in a direction making an angle of $2\tan^{-1}(\sqrt{2})$ with x-axis having characteristics parameters : 'ratio of real amplitudes' and 'difference in phases' equal to $\sqrt{2}$ and 0, respectively.

Secondly, *1st- generalization of Intensity and inadequacy of DOP* is pointed out. A simple experiment may be accomplished to record maximum and minimum intensity of a light falling on a polarizer whose fast-axis is set along a unit polarization vector, $\hat{\varepsilon}_0$. Obviously, for a light of arbitrary state of polarization one obtains extremum intensities, $(I_{\hat{\varepsilon}_0})_{max} = I_{pol} + \frac{1}{2}I_{unpol}$, and $(I_{\hat{\varepsilon}_0})_{min} = \frac{1}{2}I_{unpol}$, where $I_{pol}$, $I_{unpol}$ stands for intensities of polarized and unpolarized light respectively. DOP, in Classical Optics, is expressed as,

$$P = \frac{I_{pol}}{I_{total}} = \frac{(I_{\hat{\varepsilon}_0})_{max} - (I_{\hat{\varepsilon}_0})_{min}}{(I_{\hat{\varepsilon}_0})_{max} + (I_{\hat{\varepsilon}_0})_{min}} \ . \tag{8}$$

For polarized light, $(I_{\hat{\varepsilon}_0})_{min} = 0$ implying DOP, P=1, and for unpolarized state, DOP, P = 0 because $(I_{\hat{\varepsilon}_0})_{max} = (I_{\hat{\varepsilon}_0})_{min}$.

A natural generalization of Eq. (8) to quantum domain can be affected by replacing intensity, $I_{\hat{\varepsilon}_0}$ by Quantum Mechanical average value of the photon number operator, $\widehat{N}_{\hat{\varepsilon}_0}$, i.e., $I_{\hat{\varepsilon}_0} \to n_{\hat{\varepsilon}_0} = \text{Tr}[\hat{\rho}\,\widehat{N}_{\hat{\varepsilon}_0}]$, where Tr stands for trace, $\hat{\rho}$ is density operator for the optical field, $\widehat{N}_{\hat{\varepsilon}_0} = \hat{a}^{\dagger}_{\hat{\varepsilon}_0}\hat{a}_{\hat{\varepsilon}_0}$, $\hat{a}^{\dagger}_{\hat{\varepsilon}_0}$ ($\hat{a}_{\hat{\varepsilon}_0}$) is the creation (annihilation) operator for the optical field-mode polarized along $\hat{\varepsilon}_0$. Hence, in quantum domain, Eq. (8) takes the form

$$P^{(I)} = \frac{(n_{\hat{\varepsilon}_0})_{max} - (n_{\hat{\varepsilon}_0})_{min}}{(n_{\hat{\varepsilon}_0})_{max} + (n_{\hat{\varepsilon}_0})_{min}} \ . \tag{9}$$

Evidently, Eq. (9) is a quantum version of the definition for DOP in Classical Optics, Eq. (8) and may be regarded as $1^{st}$-generalization of Intensity [26]. Let us verify whether Eq. (9) meets basic requirements, viz., the DOP, $P^{(I)}$ attains zero-value for unpolarized state and unit-value for perfectly polarized state.

Let us consider an amplitude-coherent phase-randomized (multiphoton) optical field propagating along positive $\hat{e}_z$-direction of which quantum state, in the transverse linear polarization basis ($\hat{e}_x, \hat{e}_y$), may be specified by density operator,

$$\hat{\rho} = \frac{1}{(2\pi A_0)^2} \iint d^2\alpha_x\, d^2\alpha_y\, \delta(|\alpha_x| - A_0)\, \delta(|\alpha_y| - A_0)|\alpha_x, \alpha_y\rangle\langle\alpha_x, \alpha_y|, \tag{10}$$

where $\delta(-)$ is a Dirac-delta function; $A_0$ is a real amplitude; $|\alpha_x, \alpha_y\rangle$ are bi-modal quadrature coherent states, $(\hat{a}_x, \hat{a}_y)(|\alpha_x, \alpha_y\rangle) = (\alpha_x, \alpha_y)\,|\alpha_x, \alpha_y\rangle$, and $(\hat{a}_x, \hat{a}_y)$ are annihilation operators for x-

and y-polarized photons respectively. Noting, $\hat{a}_{\hat{\varepsilon}_0} = \varepsilon_{0x}^* \hat{a}_x + \varepsilon_{0y}^* \hat{a}_y$, Eq. (4), one obtains intensity along $\hat{\varepsilon}_0$ as

$$n_{\hat{\varepsilon}_0} = \text{Tr}[\hat{\rho}\, \hat{N}_{\hat{\varepsilon}_0}]$$

$$= \text{Tr}[\hat{\rho}\, |\varepsilon_{0x}|^2\, \hat{a}_x^\dagger \hat{a}_x + |\varepsilon_{0y}|^2\, \hat{a}_y^\dagger \hat{a}_y + \varepsilon_{0x}\varepsilon_{0y}^*\, \hat{a}_x^\dagger \hat{a}_y + \varepsilon_{0x}^*\varepsilon_{0y}\, \hat{a}_y^\dagger \hat{a}_x],$$

$$= (|\varepsilon_{0x}|^2 + |\varepsilon_{0y}|^2) A_0^2 = A_0^2, \tag{11}$$

independent of the unit polarization vector $\hat{\varepsilon}_0$ ($= \varepsilon_{0x}\hat{e}_x + \varepsilon_{0y}\hat{e}_y$ and $|\varepsilon_{0x}|^2 + |\varepsilon_{0y}|^2 = 1$). Clearly, Eq.(11) demonstrates that for this multiphoton (amplitude-coherent phase-randomized) optical field the DOP, $P^{(I)}$ is zero suggesting it, unequivocally, to the status of unpolarized state. But this is not the truth because all its quantum statistical properties are not symmetric about the direction of propagation, $\hat{e}_z$ [8]. This apposite instance breed doubts about the definition, Eq.(9) obtained through 1$^{st}$- generalization of Intensity in Quantum Optics, which, in turn, necessitates another generalization (2$^{nd}$ –generalization of Intensity).

Thirdly, *DOP through 2$^{nd}$-generalization of Intensity* is introduced by considering those measurement events in which one of the exit channels of the polarization analyzer register no photons and taking the average intensity in the other channel. It is, therefore, proposed that instead of replacing $I_{\hat{\varepsilon}_0}$ in Eq. (8) by $n_{\hat{\varepsilon}_0} = \text{Tr}\,[\hat{\rho}\, \hat{N}_{\hat{\varepsilon}_0}]$ for accomplishing the quantum version, $P^{(I)}$, Eq. (9), of DOP in Quantum Optics, we must replace $I_{\hat{\varepsilon}_0}$ by,

$$\mathcal{n}_{\hat{\varepsilon}_0} = \text{Tr}\,[\hat{\rho}\, \hat{N}_{\hat{\varepsilon}_0}\, \hat{\mathcal{V}}_{\hat{\varepsilon}_{0\perp}}], \tag{12}$$

where $\hat{\mathcal{V}}_{\hat{\varepsilon}_{0\perp}}$ is $\hat{\varepsilon}_{0\perp}$- mode's vacuum projection operator, i.e., $\hat{\mathcal{V}}_{\hat{\varepsilon}_{0\perp}} = |0\rangle_{\hat{\varepsilon}_{0\perp}} {}_{\hat{\varepsilon}_{0\perp}}\langle 0|$. Eq. (12) may be regarded as 2$^{nd}$-generalization of Intensity which leads to 2$^{nd}$ modification in DOP in Quantum Optics as,

$$P^{(II)} = \frac{(n_{\hat{\varepsilon}_0})_{\max} - (n_{\hat{\varepsilon}_0})_{\min}}{(n_{\hat{\varepsilon}_0})_{\max} + (n_{\hat{\varepsilon}_0})_{\min}}. \qquad (13)$$

We first show that the definition, Eq. (13) meets the basic requirements, i.e., the DOP, $P^{(II)}$ picks zero-value for unpolarized state and unity-value for perfectly polarized state. For a beam polarized along $\hat{\varepsilon}_0$, propagating in the $\hat{e}_z$-axis, we note $(n_{\hat{\varepsilon}_0})_{\max} = n_{\hat{\varepsilon}_0}$ and $(n_{\hat{\varepsilon}_0})_{\min} = n_{\hat{\varepsilon}_{0\perp}} = 0$ which provides unit-value for DOP, $P^{(II)}$. Next, for unpolarized light having density operator [6] in the basis $(\hat{\varepsilon}_0, \hat{\varepsilon}_{0\perp})$, $\rho = \sum_{n=0}^{\infty} B_n \sum_{r=0}^{\infty} |r, n-r\rangle_{(\hat{\varepsilon}_0, \hat{\varepsilon}_{0\perp})} {}_{(\hat{\varepsilon}_0, \hat{\varepsilon}_{0\perp})}\langle r, n-r|$, gives number of photons, $n_{\hat{\varepsilon}_0} = \sum n\, B_n = B_1 + 2B_2 + 3B_3 + \ldots$, from Eq. (12), showing independence on polarization vector $\hat{\varepsilon}_0$, which, clearly, gives value zero for DOP, $P^{(II)}$ from Eq.(13).

Moreover, correspondence of Eq. (13) can be seen through vacuum projection operator, $\mathcal{V}_{\hat{\varepsilon}_{0\perp}}$ in Weyl representation [27], i.e., $\hat{\mathcal{V}}_{\hat{\varepsilon}_{0\perp}} = (1 - :\widehat{N}_{\hat{\varepsilon}_{0\perp}}: + \frac{1}{2!}:\widehat{N}^2_{\hat{\varepsilon}_{0\perp}}: - \frac{1}{3!}:\widehat{N}^3_{\hat{\varepsilon}_{0\perp}}: + \ldots)$, where $\widehat{N}_{\hat{\varepsilon}_{0\perp}} = \hat{a}^{\dagger}_{\hat{\varepsilon}_{0\perp}} \hat{a}_{\hat{\varepsilon}_{0\perp}}$ is the number operator of virtual photons and the symbol : : denotes the normal ordering of creation and annihilation operators. Evidently, it demonstrates that if occupancy in $\hat{\varepsilon}_{0\perp}$- mode is extremely feeble, i.e. $n_{\hat{\varepsilon}_{0\perp}} \ll 1$, $\hat{\mathcal{V}}_{\hat{\varepsilon}_{0\perp}} \approx \mathbb{I}$, an identity operator, which, in turn, ensures the equality of photons in two definitions, $n_{\hat{\varepsilon}_0} \approx \mathrm{n}_{\hat{\varepsilon}_0}$ and, hence, the two definitions coincide. Also, if the beam is very weak i.e., $\mathrm{Tr}[\rho\,(\widehat{N}_{\hat{\varepsilon}_0} + \widehat{N}_{\hat{\varepsilon}_{0\perp}})] = n_{\hat{\varepsilon}_0} + n_{\hat{\varepsilon}_{0\perp}} \ll 1$, which is the case of a thermal light, the two definitions agree.

Foregoing discussion demonstrates that if the photon-numbers in orthogonal mode becomes significant, the two definitions of DOP will substantially dissimilar. Furthermore, nonlinear dependence of photons, $n_{\hat{\varepsilon}_0}$ on the orthogonally polarized photons, $n_{\hat{\varepsilon}_{0\perp}}$ leads to an interesting feature that the DOP may not only depend on the nature of the optical field, but also on the average photon-numbers (intensity). That is, if we find DOP for the field, $\rho = \int d^2\alpha \, d^2\beta \, P(\alpha,\beta)|\alpha,\beta\rangle\langle\alpha,\beta|$, and, then, increase the average photon-numbers by a factor of m without affecting the nature of the field, i.e., replacing its density operator by $\rho = \int d^2\alpha \, d^2\beta \, P(\alpha,\beta)|\sqrt{m}\,\alpha,\sqrt{m}\,\beta\rangle\langle\sqrt{m}\,\alpha,\sqrt{m}\,\beta|$, or by $\rho = \int d^2\alpha \, d^2\beta \, P'(\alpha,\beta)|\alpha,\beta\rangle\langle\alpha,\beta|$, with $P'(\alpha,\beta) = (1/m) P(\alpha/\sqrt{m}, \beta/\sqrt{m})$, the degree of polarization changes. We shall explore this peculiar aspect for two multiphoton optical fields (amplitude-coherent phase-randomized and hidden optical-polarized field) in the following discussion. Such intensity-dependent feature of various DOPs has been intensively surveyed in Ref.[10].

Finally, we shall test the efficacy of Eq. (13) to assess the polarization-states of some typical multiphoton optical fields for which the conventional definition, Eq. (9) fails as it assigns the status of unpolarized state. Applying Eq. (12) for evaluation of photon-numbers (intensity) in the mode($\hat{\varepsilon}_0, \vec{k} = k\hat{e}_z$), one obtains,

$$n_{\hat{\varepsilon}_0} = \text{Tr}[\hat{\rho}\, \hat{N}_{\hat{\varepsilon}_0}\, \hat{\mathcal{V}}_{\hat{\varepsilon}_{0\perp}}],$$

$$= \text{Tr}[\hat{\rho} \sum_{n=0}^{\infty} n \, |n, 0\rangle_{(\hat{\varepsilon}_0,\hat{\varepsilon}_{0\perp})} {}_{(\hat{\varepsilon}_0,\hat{\varepsilon}_{0\perp})}\langle n, 0|],$$

$$= \sum_{n=0}^{\infty} n \, \langle n, 0|\hat{\rho}|n, 0\rangle_{(\hat{\varepsilon}_0,\hat{\varepsilon}_{0\perp})}, \qquad (14)$$

where $\hat{\mathcal{V}}_{\hat{\varepsilon}_{0\perp}} = \sum_{n=0}^{\infty} |n, 0\rangle_{(\hat{\varepsilon}_0, \hat{\varepsilon}_{0\perp})} {}_{(\hat{\varepsilon}_0, \hat{\varepsilon}_{0\perp})}\langle n, 0|$ has been inserted. Insertion of Eq. (10) in Eq. (14), after parametrizing the basis-vectors, $\hat{\varepsilon}_0 = \cos\frac{\chi}{2} e^{-i\Delta/2} \hat{e}_x + \sin\frac{\chi}{2} e^{-i\Delta/2} \hat{e}_y$, $\hat{\varepsilon}_{0\perp} = \sin\frac{\chi}{2} e^{-i\Delta/2} \hat{e}_x - \cos\frac{\chi}{2} e^{-i\Delta/2}\hat{e}_y$, and expressing the complex amplitudes, $\alpha_{x,y} = |\alpha_{x,y}| e^{i\varphi_{x,y}}$ in polar form, yields,

$$n_{\hat{\varepsilon}_0} = (2\pi)^{-2} \int_{\varphi_x=0}^{2\pi} d\varphi_x \int_{\varphi_y=0}^{2\pi} d\varphi_y \ |\alpha_{\hat{\varepsilon}_0}|^2 \exp(-|\alpha_{\hat{\varepsilon}_{0\perp}}|^2), \tag{15}$$

with $\alpha_{\hat{\varepsilon}_0} = a\{\cos\frac{\chi}{2} \exp[i(\varphi_x + \frac{1}{2}\Delta)] + \sin\frac{\chi}{2} \exp[i(\varphi_y - \frac{1}{2}\Delta)]\}$, $\alpha_{\hat{\varepsilon}_{0\perp}} = a\{\sin\frac{\chi}{2} \exp[i(\varphi_x + \frac{1}{2}\Delta)] - \cos\frac{\chi}{2} \exp[i(\varphi_y - \frac{1}{2}\Delta)]\}$. Since the integrand in Eq. (15) involves only the difference, $\theta = \varphi_x - \varphi_y$ and not $\varphi_x$ and $\varphi_y$ independently, one may simplify Eq. (15) to yield,

$$n_{\hat{\varepsilon}_0} = (A_0^2)^2 e^{-A_0^2} \int_{\theta=0}^{2\pi} \frac{d\theta}{2\pi} [1 + \sin\frac{\chi}{2} \cos(\theta + \Delta)] \exp[A_0^2 \sin\chi \cos(\theta + \Delta)]. \tag{16}$$

Use of standard formula for modified Bessel function of order m [28], $I_m(z) = \frac{1}{\pi} \int_0^\pi d\theta \cos(m\theta) \exp[z \cos\theta]$, in the above expressions results as,

$$n_{\hat{\varepsilon}_0} = n_0 e^{-n_0} [I_0(n_0 \sin\chi) + \sin\chi \, I_1(n_0 \sin\chi)], \tag{17}$$

with intensity, $n_0 \equiv A_0^2$. Since the modified Bessel function, $I_m(x)$ is monotonically increasing function of x, for given m, the maximum and minimum intensities in $(\hat{\varepsilon}_0, \vec{k} = k\hat{e}_z)$ modes will be,

$$(n_{\hat{\varepsilon}_0})_{max} = n_0 e^{-n_0} [I_0(n_0) + I_1(n_0)], \text{ for } \chi = \frac{\pi}{2}, \tag{18}$$

$$(n_{\hat{\varepsilon}_0})_{min} = n_0 e^{-n_0}, \text{ for } \chi = 0 \text{ or } \pi, \tag{19}$$

and the value of $\Delta$ does not matter. Substituting Eqs. (18, 19) into Eq.(13) one gets,

$$P^{(II)} = \frac{(n_{\hat{\varepsilon}_0})_{max} - (n_{\hat{\varepsilon}_0})_{min}}{(n_{\hat{\varepsilon}_0})_{max} + (n_{\hat{\varepsilon}_0})_{min}} = \frac{I(n_0) - 1}{I(n_0) + 1}, \qquad (20)$$

where $I(n_0) \equiv I_0(n_0) + I_1(n_0)$. The DOP, Eq. (20), obtained through $2^{nd}$-generalization of Intensity, evidently, demonstrates the dependence on average photons ($n_0$). On limiting cases for average photons, when $n_0 \to 0$ (few photonic regime), $I(n_0) \to 1$ and $P^{(II)} \to 0$ which is palpable because for small, $n_0$ only the second-order correlation functions (Stokes theory) prevails ensuring unpolarized state. But, for intense multiphoton optical fields, i. e., $n_0 \to \infty$, $I(n_0) \to \infty$ which signifies the typical nature, $P^{(II)} \to 1$, ascertain perfect polarization.

Next, let us apply Eq. (13) for unimodular Hidden optical-polarized field [29]. It is an optical field whose characteristic polarization-parameters : 'ratio of real amplitudes' and 'sum of phases' are unity and zero, respectively, possessing non-random nature in contrast to usual polarized optical field in which 'ratio of real amplitudes and difference of phases' are non-random characteristic polarization parameters. The state of such an optical field is specified by density operator,

$$\hat{\rho} = \frac{1}{(2\pi A_0)^2} \iint d^2\alpha_x \, d^2\alpha_y \, \delta(|\alpha_x| - A_0)\delta(|\alpha_y| - A_0)||\alpha_x|e^{i\theta_x}, |\alpha_y|e^{i\theta_y}\rangle\langle|\alpha_x|e^{i\theta_x}, |\alpha_y|e^{i\theta_y}|, \quad (21)$$

with condition $\theta_x = -\theta_y = \theta^{/}$. It may be noted that the calculations for various terms in Eq. (13) proceed in similar fashions yielding same results. The Eq. (17) would be same in both cases, which leads to equivalent expression for DOP, Eq. (20) and similar interpretational tenet as that in earlier case.

*Concluding,* an expression for degree of polarization (DOP) in Quantum Optics is proposed by inserting vacuum-mode projection operator in the definition of intensity of optical field. Its correspondence with usual definition of DOP in quantum domain, derived by replacing intensity in Classical definition of DOP by quantum mechanical average values of number operators, is sought. The efficacy of the proposed definition is demonstrated for typical multiphoton optical fields where usual definition fails to predict true polarization nature. Precisely, the proposed definition of DOP uses a pragmatic approach through modifying the very definition of intensity rather than to rely on the 'abstract notion of distance' of quantum states from unpolarized state as well as to 'incomplete' correlation-functions of Stokes-variables (parameters). It is, therefore, the polarization of optical quantum field is either characterized by the criterion, Eq. (7) for perfect polarization or by the definition, Eq. (13) for partial (mixed) polarization. Moreover, for multi-modal multiphoton optical quantum fields generalization of DOP is straightforward [30].

**Acknowledgements**: We acknowledge fruitful discussions with Professors N. Chandra and R. Prakash, University of Allahabad, Allahabad, India. One of the authors (RSS) owes to Prof. Vijay A. Singh, HBCSE, Tata Institute of Fundamental Research, Mumbai, India, and Prof. Surendra Singh, Department of Physics, University of Arkansas, Fayetteville, AR (USA) for invoking inspiring comments.

[#]yesora27@gmail.com , also at: [†]Indian Institute of Information Technology, Allahabad, India

are coherent states defined by $(\hat{a}_x, \hat{a}_y) |\alpha_x, \alpha_y\rangle = (\alpha_x, \alpha_y) |\alpha_x, \alpha_y\rangle$, and $\hat{a}_{x,y}$ are annihilation operators for photons linearly polarized along x-and y- axes respectively, and propagating along z-direction, authors [6] obtained a statistical property (square of the intensity), $\text{Tr}[\hat{\rho}\,\hat{a}_\theta^{\dagger 2}\hat{a}_\theta^2] = \frac{A_0^4}{4}[5-\cos 4\theta]$, provided that $\hat{a}_\theta \equiv \hat{a}_x \cos\theta + \hat{a}_y \sin\theta$, which renders asymmetrical transverse (polarization) structure about direction of propagation. But, on account of random nature of phases of complex amplitudes, $\alpha_{x,y}$, Stokes parameters have vanishing values assigning, paradoxically, it to be in unpolarized state.